\documentclass[twocolumn,showpacs,preprintnumbers,amsmath,amssymb]{revtex4}

\usepackage{graphicx}
\usepackage{dcolumn}
\usepackage{bm}

\def\beq{\begin{equation}}
\def\eeq{\end{equation}}
\def\bw{\begin{widetext}}
\def\ew{\end{widetext}}
\def\pl{\partial}

\def\al{\alpha}
\def\bt{\beta}

\def\ga{\gamma}
\def\de{\delta}
\def\De{\Delta}

\def\Si{\Sigma}
\def\te{\theta}

\def\lam{\lambda}

\def\ep{\epsilon}

\def\sq{\sqrt}

\def\l{\left (}
\def\r{\right )}
\def\fr{\frac}
\def\la{\label}
\def\hs{\hspace}
\def\vs{\vspace}

\def\ran{\rangle}
\def\lan{\langle}
\def\ov{\overline}

\def\tm{\times}

\begin{document}


\title{Neutrino Democracy, Fermion Mass Hierarchies\\

And Proton Decay From 5D SU(5)
}

\author{Qaisar Shafi$^a$ and
Zurab Tavartkiladze$^{b, c}$ 
}

\address{{\it $^a$ Bartol Research Institute, University of Delaware,
Newark, DE 19716, USA\\
$^b$Institute for Theoretical Physics, Heidelberg University,
Philosophenweg 16, D-69120 Heidelberg, Germany\\
$^c$Institute of Physics, Georgian Academy of Sciences, Tbilisi 380077,
Georgia }}

\begin{abstract}
\vspace{0.3cm}

The explanation of various observed phenomena such as large angle neutrino
oscillations, hierarchies of charged fermion masses and CKM mixings, and
apparent baryon number conservation may have a common origin. 
We show how this could occur in 5D SUSY $SU(5)$
supplemented by a ${\cal U}(1)$ flavor symmetry and additional matter
supermultiplets called 'copies'. 
In addition, the proton decays into $p\to K\nu $, with an estimated
lifetime of order $10^{33}-10^{36}$ yrs. Other decay channels include 
$Ke$ and $K\mu $ with comparable rates.
We also expect that
BR$(\mu \to e\gamma )\sim $BR$(\tau \to \mu \gamma )$.

\end{abstract}


\begin{flushleft} 
BA-02-35 
\end{flushleft}  

\maketitle

The neutrino sector of electroweak interactions is one of the windows
which sheds light on physics beyond the standard
model. The SuperKamiokande (SK) experiments provide credible evidence for 
atmospheric
\cite{atm} and solar \cite{sol} neutrino oscillations. 
The atmospheric neutrino anomaly suggests 
$\nu_{\mu }\to \nu_{\tau }$ oscillations with 
$\De m^2_{\rm atm}\simeq 2\cdot 10^{-3}~{\rm eV}^2$ and nearly maximal
mixing angle $\sin^2 2\te_{\mu \tau}\simeq 1$.
For solar neutrinos the preferred solution seems to be the large mixing
angle MSW (LAMSW) one with  parameters
$\De m^2_{\rm sol}\simeq 6\cdot 10^{-5}~{\rm eV}^2$, 
$\sin^2 2\te_{e \mu, \tau }\approx 0.8$.
In contrast to this, the quark CKM mixing angles are small and there is
noticeable
intergeneration hierarchies between the charged fermion masses. 
It is tempting to think that there is some underlying framework
responsible for the generation of fermion masses and mixings.
A particularly attractive possibility is an abelian ${\cal U}(1)$
flavor symmetry \cite{feru1} which can be quite effective in
the charged fermion \cite{1feru1} and neutrino \cite{nuu1} sectors.

With a ${\cal U}(1)$ (or for that matter any flavor) symmetry perhaps the
most intriguing
and challenging task is to understand the origin of large (for 
$\nu_{\mu }-\nu_{\tau }$ even maximal!) neutrino mixings versus the small
CKM mixing angles. One promising scheme
is the {\it democratic approach} to neutrinos \cite{dem}, \cite{ourdem} in
which the left
handed lepton doublets $l$ are not distinguished by ${\cal U}(1)$
and consequently can mix strongly. In contrast to $l$, the quarks and
right handed leptons have distinct ${\cal U}(1)$ charges so that the
mass
hierarchies between them can be realized.

In a recent work \cite{ourdem} we examined the democratic approach
within MSSM and the GUT framework. It was shown that it is
difficult to realize neutrino 'democracy' in $SU(5)$ in a straightforward
way, whereas an
extended version of flipped $SU(5)$ allows such an implementation.
The difficulties faced in $SU(5)$ mainly arise because the 
$(10+\bar 5)_{\al }$ ($\al =1, 2, 3$ is a generation index) multiplets
unify the quark and lepton fields, leaving us with less freedom in their
${\cal U}(1)$ charge assignments.
To overcome this, one can think of some reasonable
extension in such a way as to 'split' the fermion fragments of GUT
multiplets
and thereby relax the unwanted constraints on ${\cal U}(1)$ charges. This
may
not be a trivial task in $SU(5)$ from one's earlier experience with the
doublet-triplet (DT) splitting problem in the scalar sector
which ends up
requiring a rather complicated extensions. However, these arguments and also
the discussions of ref. \cite{ourdem} are valid for four dimensional
constructions. Recent developments in higher dimensional orbifold
constructions \cite{kawa}-\cite{mu3} have shown that many outstanding
problems of GUTs
can be resolved in an extra-dimensional setting. Namely, the orbifold
setting can be exploited to yield natural DT splitting, GUT symmetry
breaking
and baryon number conservation (to the desired level).
Note that the concept of 'split' multiplets has previously appeared in the
framework of superstring theories \cite{split}.

In this letter we will apply the orbifold approach for realizing the
observed
fermion mass pattern consistent with neutrino democracy within 5D SUSY
$SU(5)$, supplemented by a ${\cal U}(1)$ flavor symmetry. 
We show how
the problems discussed above can be nicely avoided by invoking a fifth
dimension, with additional supermultiplets (so-called 'copies') playing an
essential role. The ${\cal U}(1)$ symmetry also helps ensure sufficient
proton stability.
Due to democracy in the left handed lepton sector, the decays $p\to Ke$,
$p\to K\mu $ are expected to proceed at comparable rates. 
Similarly, the branching rates in the lepton sector satisfy
BR$(\mu \to e\gamma )\sim $BR$(\tau \to \mu \gamma )$, which may be
testible in the future.

\vs{0.3cm}

Consider 5D $N=1$ SUSY $SU(5)$ compactified on an $S^{(1)}/Z_2\tm Z_2'$
orbifold, such that the 'low' energy theory has the MSSM field
content \cite{kawa}. 
In 4D $N=1$ superfield notation, the 5D gauge
supermultiplet is $V_{N=2}=(V, \Si )$, where $V$ and $\Si $
respectively are vector and chiral superfields, both in the adjoint 
${\bf 24}$ representation of $SU(5)$. In terms of 
$SU(3)_c\tm SU(2)_L\tm U(1)_Y\equiv G_{321}$, 
$V(24)=V_c(8, 1)_0+V_w(1, 3)_0+V_s(1, 1)_0+
V_X(3, \bar 2)_5+V_Y(\bar 3, 2)_{-5}$, where the subscripts denote
the
hypercharge $Y=\fr{1}{\sq{60}}{\rm diag}(2, 2, 2, -3, -3)$ in the units of
$1/\sq{60}$. Similar decomposition holds for $\Si (24)$.
It is assumed that the fifth space like dimension $y$ describes a compact 
$S^{(1)}$ circle with radius $R$. Under the $Z_2\tm Z_2'$ symmetry 
$Z_2$: $y\to -y$, $Z_2'$: $y'\to -y'$ ($y'=y+\pi R/2$), all 
states should have definite $Z_2\tm Z_2'$ parities $(P, P')$.  
With the following parity assignments for the fragments from 
$V_{N=2}(24)$
$$
\l V_c,~V_w,~V_s\r \sim (+,~+)~,~~~~ 
\l V_X,~V_Y\r \sim (-,~+)~,  
$$
\vs{-0.8cm}
\beq
\l \Si_c,~\Si_w,~\Si_s\r \sim (-,~-)~,~~~~ 
\l \Si_X,~\Si_Y\r \sim (+,~-)~,
\la{gaugepar}
\eeq
on the fixed point $y=0$ (identified with our 4D world) we
will have $N=1$
SUSY with massless $G_{321}$ gauge bosons. The remaining states in
$SU(5)/G_{321}$ acquire
large (GUT scale) masses. Thus, states with parities $(\pm , \pm)$,
$(\pm , \mp )$ respectively have masses $(2n+2)/R$, $(2n+1)/R$, where $n$
denotes the quantum number in the KK mode expansion.

The $SU(5)$ 'higgs' superfields, which contain the pair of MSSM higgs
doublets, are also introduced in the bulk. Namely, there are two $N=2$ 
supermultiplets ${\cal H}_{N=2}(5)=(H, \ov{H}\hs{0.2mm}')$,
$\ov{\cal H}_{N=2}(\bar 5)=(\ov{H}, H')$, where
$H$, $\ov{H}$ are $5$, $\bar 5$ plets of $SU(5)$ and
$\ov{H}\hs{0.2mm}'$, $H'$ respectively are their mirrors. In terms of
$G_{321}$,
$H(5)=h_u(1, 2)_{-3}+T(3, 1)_2$,
$\ov{H}(\bar 5)=h_d(1, \bar 2)_3+\ov{T}(\bar 3, 1)_{-2}$ and similarly for
$H'$, $\ov{H}\hs{0.2mm}'$. With $Z_2\tm Z_2'$ parity assignments
$$
(h_u,~h_d)\sim (+,~+)~,~~({h_d}',~{h_u}')\sim (-,~-)~,
$$
\vs{-0.8cm} 
\beq
(T,~\ov{T})\sim (-,~+)~,~~(\ov{T}\hs{0.3mm}',~ T')\sim (+,~-)~, 
\la{higgspar}
\eeq
only $h_u$, $h_d$ have zero modes and can be identified with the MSSM
doublets. All colored triplet partners become superheavy and in this way
the DT splitting occurs naturally. Note that since 
$h_u$, $h_d$ arise from different $N=2$ supermultiplets, the mass
term $M_hh_uh_d$ is not allowed in 5D. This can be considered a
good starting point for obtaining an adequately suppressed $\mu $-term
(however, at 4D level additional care must be exercised \cite{mu2},
\cite{mu3} for avoiding a large $\mu $-term).

In orbifold constructions with a
minimal setting, the introduction of fermions
in the bulk is not straightforward. For example, the $d^c$ and $l$
states,
which come from the same $\bar 5$ plet, are involved in a 
5D kinetic coupling $l^+V_Xd^c$ and, since the $V_X$ boson has parity 
$(-, +)$
[see (\ref{gaugepar})], either $d^c$ or $l$ should have parity $(-,
+)$. This would mean loss of some (zero mode) MSSM chiral
states. For overcoming this difficulty, one could attempt to introduce
chiral states not in the bulk but directly on a brane with no
KK
excitations. Although the theory would be fully selfconsistent,
we follow here a different procedure and introduce in the bulk
additional
supermultiplets \cite{arch}  called 'copies' 
(see $1^{\rm st}$ and $3^{\rm rd}$ refs. in \cite{symbr}), denoted by
$10'+\bar 5'$ (per generation). By suitable prescription of $Z_2\tm Z_2'$
parities, these states allow one to realize at low energies complete three
generations of MSSM
massless chiral states. It will turn out that the introduction of 
these copies enables us
to realize the democratic approach for neutrinos and obtain a nice picture
for the charged fermion sector. From this point of view, the motivation
for
introducing copies therefore becomes twofold.

In the bulk we introduce three generations of $N=2$
supermultiplets
${\bf {\cal X}}_{N=2}=(10,~ \ov{10})$, 
${\bf \ov{\cal V}}_{N=2}=(\bar 5,~5)$ together with their copies
${{\bf {\cal X}}\hs{0.3mm}'}_{N=2}$, 
${{\bf \ov{\cal V}}\hs{0.3mm}'}_{N=2}$. Recall that in terms of $G_{321}$
$10=e^c(1, 1)_{-6}+q (3, 2)_{-1}+u^c (\bar 3, 1)_{4}$, 
$\bar 5=l (1, \bar 2)_3+d^c (1, \bar 3)_{-2}$
and likewise for $10'$, $\bar 5'$. The $\ov{10}$, $5$, $\ov{10}'$, $5'$
are mirrors and their fragments have conjugate transformation properties
under $G_{321}$.
The following orbifold parity assignments
$$
\l q',~l,~{u^c},~{d^c}', ~{e^c} \r \sim (+,~+)~,
$$
\vs{-0.8cm}  
\beq
\l q,~l',~{u^c}',~{d^c}, ~{e^c}' \r \sim (-,~+)~,  
\la{ferpar}  
\eeq 
with opposite parities for the corresponding mirrors (we assume generation
independent parities), are consistent with the prescriptions in
(\ref{gaugepar}), and it is easy to verify that all $N=2$ SUSY invariant
terms also possess $Z_2\tm Z_2'$ invariance.
From (\ref{ferpar}) we see that the states
$e^c, q', u^c, l, {d^c}'$ contain
zero modes which we identify with the three chiral quark-lepton families
of
MSSM.

In addition, we introduce a ${\cal U}(1)$ flavor symmetry
(on whose origin we will comment later) and a singlet superfield $X$
carrying
${\cal U}(1)$ charge $Q_X=-1$. We assume 
$\fr{\lan X\ran}{M}\equiv \ep \simeq 0.2$ ($M$ is some cut off
close to the fundamental scale). Because of the fact
that the 'matter'
states come from different $SU(5)$ multiplets, the
constraints on ${\cal U}(1)$ charge assignments are more relaxed (this
turns out to be sufficient to obtain a nice and consistent picture).
We have only one constraint
\beq
Q[{u^c}_{\al }]=Q[{e^c}_{\al }]~.
\la{con}
\eeq
The ${\cal U}(1)$ charges for matter states are chosen as follows:
$$
Q[{d^c_1}']=b+c-a+k+2\hs{0.3mm},~~
Q[{d^c_2}']=Q[{d^c_3}']=b+c-a+k\hs{0.3mm}, 
$$
\vs{-0.8cm}  
$$
Q[u^c_1]=Q[e^c_1]=b+5\hs{0.3mm},~Q[u^c_2]=Q[e^c_2]=b+2\hs{0.3mm},~
$$
\vs{-0.8cm}  
$$
Q[u^c_3]=Q[e^c_3]=b\hs{0.3mm},~~Q[l_1]=Q[l_2]=Q[l_3]=c+k\hs{0.3mm},
$$
\vs{-0.8cm}  
\beq
Q[q_1']=a+3\hs{0.3mm},~Q[q_2']=a+2\hs{0.3mm},~Q[q_3']=a\hs{0.3mm},~  
\la{charges}
\eeq
($k\stackrel{>}{_-}0$ is an integer and $a$, $b$, $c$ are some phases 
undetermined for the time being). Note that (\ref{con}) is
satisfied for each generation and $l_{\al }$ all have the same ${\cal 
U}(1)$ charge.
Assuming $Q(h_u)=-a-b$, $Q(h_d)=-b-c$, the relevant couplings
generating the up and down quark and charged lepton masses respectively
are
\begin{equation}
\begin{array}{ccc}
 & {\begin{array}{ccc}
\hspace{-5mm} u^c_1 & \,\,~~u^c_2 ~ & \,\,u^c_3 ~
\end{array}}\\ \vspace{2mm}
\begin{array}{c}
q_1' \\ q_2' \\q_3'
 \end{array}\!\!\!\!\! &{\left(\begin{array}{ccc}
\,\,\epsilon^8~~ &\,\,\epsilon^5~~ &
\,\,\epsilon^3  
\\  
\,\,\epsilon^7~~   &\,\,\epsilon^4~~  &
\,\,\ep^2
 \\
\,\,\epsilon^5~~ &\,\,\epsilon^2 ~~ &\,\,1
\end{array}\right)\hs{-0.5mm}h_u }\hs{0.3mm}, 
\end{array}  \!\!  ~~
\begin{array}{ccc}
 & {\begin{array}{ccc}
\hspace{-5mm} ~~~~{d^c_1}'~ & \,\,{d^c_2}' ~~ & \,\,{d^c_3}' ~~~~~~

\end{array}}\\ \vspace{2mm}
\begin{array}{c}  
q_1' \\ q_2' \\q_3'
 \end{array}\!\!\!\!\! &{\left(\begin{array}{ccc} 
\,\,\epsilon^5~~ &\,\,\epsilon^3~~ &
\,\,\epsilon^3
\\
\,\,\epsilon^4~~   &\,\,\epsilon^2~~  &
\,\,\epsilon^2
 \\
\,\,\epsilon^2~~ &\,\,1~~ &\,\,1
\end{array}\right)\hs{-0.1cm}\epsilon^kh_d}\hs{0.3mm},
\end{array}  \!\!  ~~~~~
\label{updown}
\eeq
\vs{-0.8cm}
\begin{equation}
\begin{array}{ccc}
 & {\begin{array}{ccc}
\hspace{-7mm} e^c_1~~ & \,\,e^c_2 ~ & \,\,e^c_3 
  
\end{array}}\\ \vspace{2mm}
\begin{array}{c}
l_1 \\ l_2 \\l_3
 \end{array}\!\!\!\!\! &{\left(\begin{array}{ccc}
\,\,\epsilon^5~~ &\,\,\epsilon^{2}~~ &
\,\,1
\\  
\,\,\epsilon^5~~   &\,\,\epsilon^2~~  &
\,\,1
 \\
\,\,\epsilon^5~~ &\,\,\epsilon^2~~ &\,\,1
\end{array}\right)\epsilon^kh_d }~.
\end{array}  \!\!  ~~~~~
\label{lept}
\eeq
The entries in textures (\ref{updown}) and (\ref{lept})
are taken for simplicity to be
real and are accompanied by factors of order unity (here we
will not concern ourselves with CP violating phases).
Diagonalization of (\ref{updown}), (\ref{lept}) yields for the Yukawa
couplings
\beq
\lambda_t\sim 1~,~~
\lambda_u :\lambda_c :\lambda_t \sim
\epsilon^8:\epsilon^4 :1~,
\la{ulam}
\eeq
\vs{-0.8cm}  
\beq
\lam_b\sim \lam_{\tau}\sim \ep^k ~,~~~
\lambda_d :\lambda_s :\lambda_b \sim
\epsilon^5:\epsilon^2 :1~,
\label{dlam}
\eeq
\vs{-0.8cm}
\beq
\lambda_e :\lambda_{\mu } :\lambda_{\tau } \sim
\epsilon^5:\epsilon^2 :1~,
\label{elam}
\eeq
which have the desired hierarchical pattern.
{}From (\ref{updown}), we find
\beq
V_{us}\sim \epsilon~,~~~V_{cb}\sim \epsilon^2~,~~~
V_{ub}\sim \epsilon^3~,
\label{ckm}
\eeq
values that are consistent with the observations.

{}From (\ref{lept}) the expected values for the lepton mixing angles are
\beq
\sin^2 2\te_{\mu \tau }\sim 1~,~~~\sin^2 2\te_{e \mu, \tau }\sim 1~,
\la{lepmix}
\eeq
which nicely fit with the SK data. To generate neutrino masses we
introduce two right handed neutrinos
$N$, $N'$ [in 5D they are accompanied by appropriate mirrors
$\ov{N}$, $\ov{N}\hs{0.3mm}'$ with parities $(-, -)$] 
with ${\cal U}(1)$ charges
$Q(N)=p+1/2$, $Q(N')=q+1/2$ ($p, q$ are positive integers). With
\beq
a+b=\fr{1}{2}~,~~~~c=0~,
\la{select}
\eeq
and taking into account (\ref{charges})
the relevant couplings are
$$
\ep^{k+p}(\lam_1l_1+\lam_2l_2+\lam_3l_3)Nh_u+\ep^{2p+1}M_NN^2+
$$
\vs{-0.8cm}
$$
\ep^{k+q}(\lam_1'l_1+\lam_2'l_2+\lam_3'l_3)N'h_u+\ep^{2q+1}M_N'{N'}^2+
$$
\vs{-0.8cm}
\beq
\ep^{p+q+1}M_{NN'}NN'~,
\la{dirmaj}
\eeq
where $\lam_{\al }$, $\lam_{\al }'$ are dimensionless coefficients of
order
unity. With $p>q$, $\ep^{2q}M_N'\gg \ep^{2p}M_{NN'}$,
$M_{NN'}^2\ll M_NM_N'$ (these assumptions are needed for the correct
scales
of lepton number violations whose origin is still unexplained in this
setting), integration of $N$, $N'$ states leads to the
neutrino mass matrix
\beq
m^{\nu }_{\al \bt}=\lam_{\al }\lam_{\bt }m+\lam_{\al }'\lam_{\bt }'m'~,
\la{numatr}
\eeq
where $m=\ep^{2k-1}h_u^2/M_N$, $m'=\ep^{2k-1}h_u^2/M_N'$. For
$M_N/\ep^{2k-1}\simeq 2\cdot 10^{14}$~GeV
and $M_N'/\ep^{2k-1}\simeq 1.2\cdot 10^{15}$~GeV we have
$m\simeq 5\cdot 10^{-2}$~eV, $m'\simeq 8\cdot 10^{-3}$~eV. Ignoring 
the subleading
term in (\ref{numatr}), one finds that only $m_{\nu_3}$ acquires a
mass $m_{\nu_3}=(\lam_1^2+\lam_2^2+\lam_3^2)m$, while in this limit 
$m_{\nu_1}=m_{\nu_2}=0$. Therefore, 
$\De m_{\rm atm}^2\simeq m_{\nu_3}^3\sim 10^{-3}~{\rm eV}^2$. 
The second term in
(\ref{numatr}) gives rise to a mass $m_{\nu_2}\sim m'$, so that 
$\De m_{sol}^2\sim {m'}^2\simeq 6\cdot 10^{-5}~{\rm eV}^2$, 
the scale relevant for LAMSW solution. We therefore conclude that the
desirable $\nu_e\to \nu_{\mu, \tau }$, $\nu_{\mu }\to \nu_{\tau }$
oscillation scenarios are realized within our 5D framework.

\vs{0.1cm}

Let us now turn to the issue of baryon number and matter parity
violation. With the selections (\ref{charges}), (\ref{select}), it is easy
to
verify that matter parity violating operators $lh_u$, $e^cll$,
$q'l{d^c}'$,
$u^c{d^c}'{d^c}'$ are forbidden to all orders if $a$ is either an integer,
or $a>2k+1/2$. The ${\cal U}(1)$ symmetry can also
forbid dimension three and four baryon number violating
operators. As fas as the $d=5$ operators are concerned, because of the
absence of zero modes in colored triplet 'scalar' states, 
potential nucleon decay
through their exchange does not arise in orbifold SUSY $SU(5)$. The
non-renormalizable $d=5$ operators $q'q'q'l$ and $u^cu^c{d^c}'e^c$ are
eliminated if both $3a$ and $5a$ are non
integers (this choice is also compatible with matter parity
conservation).

Dimension five nucleon decay at measurable rates could occur
through the couplings 
$\fr{\ov{\lam }}{M}{q_1}'{q_1}'{q_2}'l_{\al }$ if
 $a=(3m+1)/3$, such that
$\ov{\lam }\sim \ep^{9+3m+k}$. For $M\sim 10M_G$ and
$(k, m)=(0, 1), (1, 1), (2, 1), (3, 0)$, we find 
$\tau (p\to K\nu_{\al })\sim 10^{33}-10^{36}$ yr
(note that $\tan \bt \sim \fr{m_t}{m_b}\ep^k$). 
The above couplings also lead to proton decay with emission of charged
leptons. However,
their rates are smaller by a factor of $\sim 10$. The democratic scenario
predicts that $p\to Ke$ and $p\to K\mu $
proceed with nearly equal rates, such that
$\tau (p\to Ke)\sim \tau (p\to K\mu )\sim 10^{34}-10^{37}$ yr.

As far as dimension $6$ nucleon decay is concerned,  with
all
matter introduced in the bulk and due to the copies, the 5D bulk kinetic
terms
are irrelevant for nucleon decay \cite{arch}.
This is because through the exchange of $V_X$, $V_Y$ bosons, the
light 
quark-lepton states are converted into heavy states with masses
of  order $1/R$. The only source for nucleon decay could be some brane
localized non-diagonal kinetic operators allowed by $G_{321}$ and orbifold
symmetries. Such operators have the form \cite{locop}
$\de (y)\psi_1^+(\pl_5e^{2\hat{V}}-\hat{\Si }e^{2\hat{V}}-
e^{2\hat{V}}\hat{\Si })\psi_2$, where $\psi_1$ and $\psi_2$ denote quark
and lepton superfields respectively, and $\hat{V}$, $\hat{\Si }$ 
are fragments
from the coset $SU(5)/G_{321}$. 
In order for these operators be invariant under ${\cal U}(1)$,
the multiplier $(X^+)^{Q_1}X^{Q_2}$ will be present, where $Q_1$, $Q_2$
are the ${\cal U}(1)$ charges of $\psi_1$, $\psi_2$. 
If either $Q_1$ or $Q_2$ is not an integer, 
the corresponding operator is not allowed.
From
(\ref{charges}), (\ref{select}) we verify that  either $Q[{q_{\al }}']$
or
$Q[u^c_{\al }]=Q[e^c_{\al }]$ is not an ineger, and consequently
the couplings
$({q'}^+e^c+{u^c}^+q')\pl_5V_X$, $({e^c}^+q'+{q'}^+u^c)\pl_5V_Y$
are absent. For $a=(2m+1)/4$ ($m$ is an integer), the couplings
$l^+{d^c}'\pl_5V_X(X^+)^kX^{m-k}$ will appear, but these terms alone do
not induce nucleon decay.
Thus, thanks to the ${\cal U}(1)$
symmetry, $d=6$ nucleon decay is absent.

%
%
\vs{0.2cm}

We conclude with some observations:

{\it a)} The leptonic mixing angles
$\te_{12}$, $\te_{23}$ receive 
contributions both from the charged lepton and neutrino
sectors. In the absence of cancellations between these contributions
we expect (\ref{lepmix}) to hold naturally. 
However, the CHOOZ data \cite{chooz} requires
$\te_{13}\stackrel{<}{_\sim }0.2(\simeq \ep)$, so that some
cancellation between contributions from the two sectors is
needed. If future measurements turn out to favor a much smaller
($\ll \ep $) 
$\te_{13}$, then some new explanation would be required.

{\it b)} The democratic approach also has  important implications for 
lepton flavor violating rare
proccesses.
Since the neutrino Dirac Yukawa
couplings in (\ref{dirmaj}) for different families are all of the same
order, one can expect that 
BR$(\mu\to e\ga )\sim$BR$(\tau \to \mu \ga) $. For universal 
(at high scale) sparticle masses$\sim 300-500$~GeV and 
$\tan \bt =25-50$,
the constraint BR$\stackrel{<}{_\sim }10^{-14}$ (the most stringent
expected bound for $\mu \to e\ga $ \cite{BR}) requires $p>q=2, 3$. 
For
$\tan \bt \sim 1-5$ we can have $p>q=0$. 


{\it c)} The origin of lepton number violation scale (masses of right
handed
neutrinos) is unexplained in this setting. This is not surprizing in
$SU(5)$, but in GUTs such as $SO(10)$ or even 
$SU(4)_c\tm SU(2)_L\tm SU(2)_R$, the violation of lepton
number is directly related to the $B-L$ breaking scale which, in
a minimal setting,
is close to $10^{16}$~GeV. Thus, it would be interesting to extend the
present discussion to such models.

{\it d)} While the ${\cal U}(1)$ flavor symmetry can provide an 
understanding  of why proton decay has so far not been seen, it remains to
be seen if dimension five operators should be expunged or not. Hopefully,
future measurements will shed more light on this fundamental question,
help determine some of the ${\cal U}(1)$ charges and test the democratic
approach by comparing decays with emission of charged leptons.

{\it e)} The ${\cal U}(1)$ flavor symmetry can be
global or even can be substituted by some discrete ${\cal Z}_N$ symmetry
which arises in the  fermion sector from some more fundamental
theory. If ${\cal U}(1)$ is introduced in 5D  as a vector-like
gauge symmetry \cite{bl}, after compactification it can cause localized
anomalies on the orbifold fixed points \cite{locan}. Their cancellation
could
occur through bulk Chern-Simons term, with possibly some additional states
playing an essential role \cite{1su3w}, \cite{mu2}, \cite{bl}.

{\it f)} One could imagine extending the ${\cal U}(1)$ flavor symmetry
to non-abelian flavor groups such as $SU(2)_H$ or
$SU(3)_H$ within the orbifold constructions. Such flavor symmetries,
apart from providing an expanation of hierarchical structures, 
may yield additional predictions
and relations between fermion masses and their mixings.

\vs{0.2cm}

We acknowledge the support of NATO Grant
PST.CLG.977666. This work is partially supported  by DOE under
contract DE-FG02-91ER40626.

\vs{-0.3cm}

\end{document}